\documentclass{JAC2000}

\usepackage{graphicx}

\begin{document}
\title{LATEST DEVELOPMENTS FROM THE S-DALINAC\thanks{Supported by
    DFG under contract no. FOR~272/2-1 and Graduiertenkolleg
    "Physik und Technik von Beschleunigern"}}

\author{\noindent M.~Brunken, H.~Genz, M.~Gopych, H.-D.~Graef, S.~Khodyachykh,
    S.~Kostial, U.~Laier, A.~Lenhardt,\\
    H.~Loos, J.~Muehl, M.~Platz, A.~Richter, S.~Richter,
    B.~Schweizer, A.~Stascheck, O.~Titze,\\
    S.~Watzlawik (TU Darmstadt), S.~Doebert (CERN)}

\maketitle

\begin{abstract}
The S-DALINAC is a 130 MeV superconducting recirculating
electron accelerator serving several nuclear and radiation physics experiments
as well as driving an infrared free-electron laser.
A system of normal conducting rf resonators for noninvasive beam
position and current measurement was established.
For the measurement of gamma-radiation inside the accelerator cave
a system of Compton-diodes has been developed and tested.
Detailed investigations of the transverse phase space were carried
out with a tomographical reconstruction method of optical transition
radiation spots. The method can be applied also to non-Gaussian
phase space distributions. The results are in good accordance with simulations.
To improve the quality factor of the superconducting 3 GHz cavities,
an external 2K testcryostat was commissioned. The influence of
electro-chemical polishing and magnetic shielding is currently under
investigation. A digital rf-feedback system for the accelerator cavities
is being developed in order to improve the energy spread of the beam
from the S-DALINAC.

\end{abstract}

\section{Introduction}

A comprehensive discussion of the layout and the properties of the
recirculating superconducting electron accelerator S-DALINAC
is given in~\cite{bib:S-DALINAC}. The electrons are emitted by a
thermionic gun and then accelerated electrostatically to 250~keV. A 
normal conducting 3~GHz chopper-prebuncher system creates the required
3~GHz time structure of the beam.
An additional subharmonic 600~MHz chopper/buncher allows for a
10~MHz bunch repetition rate
for FEL operation. The superconducting injector linac consists of one 2-cell
capture cavity ($\beta$=0.85), one 5-cell cavity ($\beta$=1), and two 20-cell
cavities operated in liquid helium at 2~K. The electron beam behind the injector
with a maximum energy of 10~MeV can either be directed to a first experimental 
site or it can be
injected into the main linac. There, eight 20-cell cavities provide 
an energy gain of up to 40~MeV.
When leaving the main linac, the beam can be extracted to the
experimental hall or it can be recirculated and reinjected one or two times.
The
maximum beam energy after three passes through the linac amounts to 130~MeV. An
infrared FEL with wavelengths between 3 and 10 $\mu$m is driven by the electron
beam with an energy from 25 up to 50~MeV.

For the different experiments, a beam current from some nA up to 60~$\mu$A can
be delivered. In the subharmonic injection mode, a peak current of 2.7~A can be
passed through the FEL undulator.

\section{Beam- and Position Monitors}
 
\begin{figure}[htb]
\centering
\includegraphics*[width=75mm]{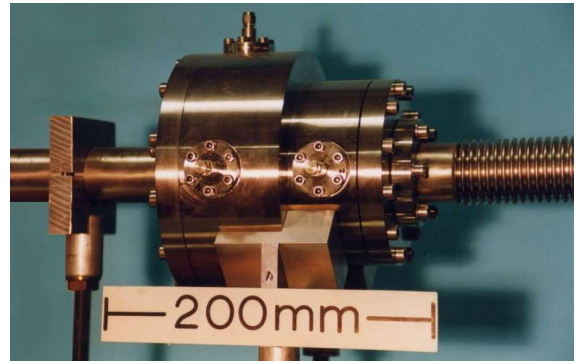}
\caption{Non intercepting 3 GHz rf monitor.}
\label{fig:monitor}
\end{figure}

A combination of normal conducting TM$_{010}$- and TM$_{110}$-cavities as
displayed in fig.~\ref{fig:monitor} was recently developed for the S-DALINAC
to measure
the beam intensity and position. The cavities are fabricated from stainless
steel, they have a common centerpiece and two covers which connect to the
beam line. The rf outputs use ceramic feedthroughs. The monitors are
operated at loaded Qs of less than 1000. Thus, they need no frequency or
temperature stabilization.

The sensitivity is 15~nW/($\mu$A)$^2$ for the intensity monitor and
15~pW/(mm~$\mu$A)$^2$ for the position monitor. For the detection of the
rather low signals, lockin techniques are used. Dedicated electronics close
to the monitor convert the signal to a dc voltage, enabling even the measurement
of a 0.1~mm beam position change at a beam current of 1~$\mu$A or a
10~nA current change. Seven monitor
units have been installed in different sections of the accelerator. The
monitor signals can be displayed graphically in the S-DALINAC control room.

\section{Compton-Diodes}

\begin{figure}[htb]
\centering
\includegraphics*[width=75mm]{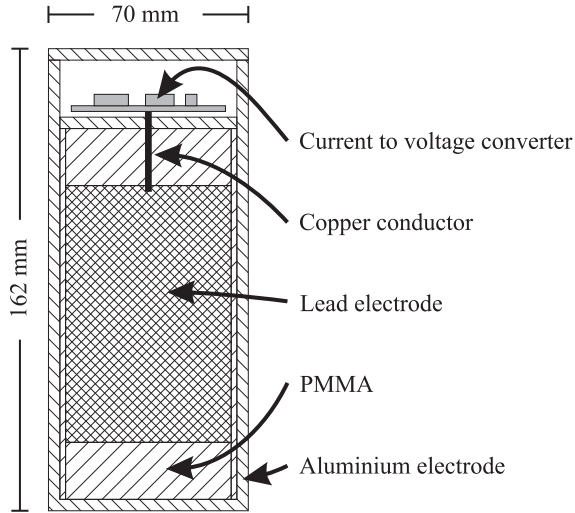}
\caption{Layout of a Compton-diode.}
\label{fig:Compton}
\end{figure}

\begin{figure}[htb]
\centering
\includegraphics*[angle=90,width=75mm]{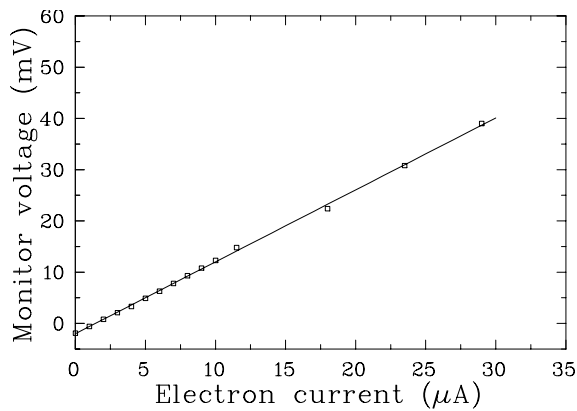}
\caption{Linearity of the Compton-diode shown in fig.~\ref{fig:Compton}.}
\label{fig:stromlinear}
\end{figure}

For a detailed examination of effects of the bremsstrahlung background in the
accelerator cave on accelerator system components, a monitoring system has
been constructed and is currently being tested. The layout of the 
bremsstrahlung monitors (also referred to as Compton-diodes) is shown in
fig.~\ref{fig:Compton}.
They consist of an inner lead electrode and
an outer aluminium electrode insulated by plexiglas. Due to the different
Compton cross sections of the electrodes, a photon beam penetrating the
monitor creates a small current between the electrodes, typically several
pA for a dose rate of 10~mSv/hr. This current is converted to a voltage,
amplified and read out via ADCs. The linearity of the output voltage over
the photon flux was demonstrated at a radiation physics setup behind
the injector. The electrons were targeted onto a copper bremsstrahlung 
converter, the resulting gamma beam was collimated by a copper collimator.
Figure~\ref{fig:stromlinear} shows the monitor output voltage as a function
of the electron current on the converter target. The Compton diodes are very
rugged and form a
flexible system which can monitor any location outside the beam pipe.
Thus radiation impact on accelerator components can be measured and beam losses
can be detected.

\section{Transverse Phase Space Tomography}
 
\begin{figure}[htb]
\centering
\includegraphics*[width=75mm]{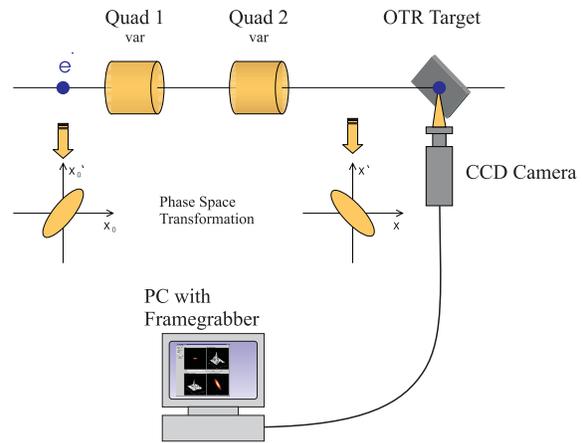}
\caption{Set up for phase space tomography.}
\label{fig:tomography}
\end{figure}

The method of transverse phase space tomography~\cite{bib:tomography}
has been applied to the
electron beam behind the injector of the S-DALINAC. The setup shown in
fig.~\ref{fig:tomography}
consists of an optical transition radiation (OTR) target, a CCD camera and
a PC with a framegrabber board. Two quadrupoles have been used to change the
beam transport matrix accordingly. A computer code written in the
Interactive Data Language (IDL) reconstructs the transverse phase space with
a tomographical algorithm. The advantage of this method is the capability of
reconstructing the phase space distribution without assuming any particular
shape.

The accuracy of the reconstruction algorithm was tested by simulations.
A total of 18 projections of a non-symmetric distribution interpolated
to 90 projections
lead to a reconstruction result with an emittance error of less than 15\%.
First measurements with an 8~MeV electron beam showed good agreement of
the so determined emittance with the one from the common method.

\section{Q-Value of the Accelerator Cavities}

The accelerator cavities used at the S-DALINAC are operated at 2~K, the
frequency of the $\pi$-mode, used for acceleration is 2.997~GHz. The design
parameters of the 1~m long 20-cell cavities consisting of niobium (RRR=280)
assumed an unloaded quality factor of 3$\cdot$10$^9$ and an accelerating
gradient of 5~MV/m.
Almost all gradients achieved during routine operation exceed this design
criteria, some resonators reach up to 10~MV/m. On the other hand,
although different preparation techniques have been tested, currently none of
the cavities has achieved a Q-value significantly higher than 1$\cdot$10$^9$.
The reduction of the Q-values in comparison with the design criteria
increases the dissipated power per cavity from 4.2 to 12.6~W. As a consequence
the maximum energy of the S-DALINAC in the cw-mode is limited by the installed
He-refrigerator power.
A measurement of the quality factor as a function of temperature has
revealed that the resonators have a residual resistance of 276~n$\Omega$ compared
to the BCS-resistance of 50~n$\Omega$ at 2~K.
  
\begin{figure}[htb]
\centering
\includegraphics*[width=75mm]{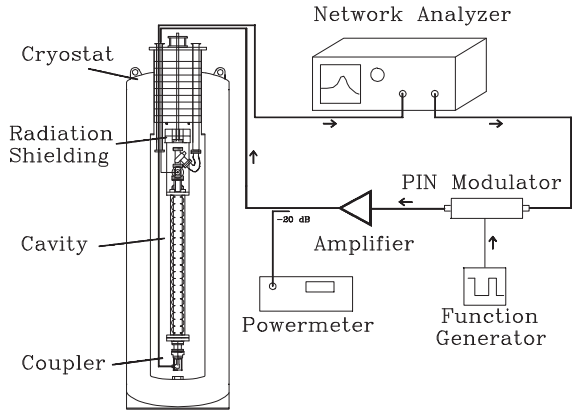}
\caption{Layout of the external 2~K testcryostat.}
\label{fig:testcryostat}
\end{figure}

In order to find an explanation for this behaviour a vertical 2~K testcryostat
was turned into operation (see fig.~\ref{fig:testcryostat}).
This test setup allows to perform systematic
studies without interfering with accelerator operation. We intend to develop an
improved magnetic shielding for the cavities which takes the constraints of the
complicated geometry (couplers, tuners) better into account than the present
shielding. Additionally,
systematic studies on the influence of different surface and material
preparation methods on the Q-value are planned.

\section{Digital RF-Control System}

\begin{figure}[htb]
\centering
\includegraphics*[width=75mm]{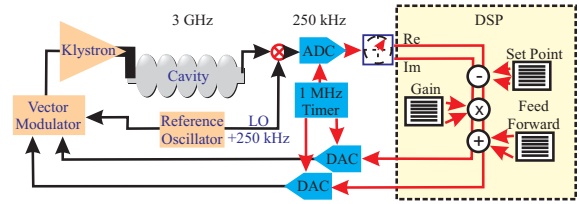}
\caption{Digital rf-control system.}
\label{fig:digitalrf}
\end{figure}

The superconducting accelerator cavities have to be controlled to an rf phase
error of less than 1$^{\circ}$ and a relative amplitude error of less than
$\pm$1$\cdot$10$^{-4}$.
The present analog control system fullfills the phase specifications, but
it does not quite meet the amplitude specifications and it
does not allow the use of modern digital control methods or detailed
control data analysis. Figure~\ref{fig:digitalrf} displays the schematic
layout of a new digital control system which is currently under development
in cooperation with DESY, Hamburg~\cite{bib:DESY}. The 3~GHz signal extracted from a sc
cavity is converted down to an intermediate frequency of 250~kHz. An ADC samples
this signal at a rate of 1~MHz yielding a complex field vector. A digital
signal processor (DSP) using techniques like feed forward tables
creates a new output field vector. This vector is converted by DACs and
mixed up to 3~GHz, amplified by klystrons and fed into the cavity.
The remaining energy spread of the electron beam should be smaller by a factor
of three with the new sytem.
  
\section{Conclusion}

At the S-DALINAC, several improvements were made with respect to beam
diagnostics. Especially the rf intensity and position monitors as well
as the Compton diodes will give substantial aid in linac operation.
The tomographical phase space reconstruction will provide more detailed
information on the electron beam structure. The studies on cavity Q-values
will hopefully result in a higher average Q, thus enabling a higher achievable
beam energy. The new digital rf system should reduce the energy spread of
the beam and improve the stability of accelerator operation.

\end{document}